\RequirePackage{fix-cm}
\documentclass[smallextended]{svjour3}       
\smartqed  
\usepackage{graphicx}
\smartqed  
\usepackage{color}
\usepackage[numbers,sort&compress]{natbib}
\usepackage[version=3]{mhchem}
\usepackage{amssymb}
\usepackage{enumerate}
\usepackage{longtable}

\usepackage{ulem}
\usepackage{graphicx}
\usepackage{xcolor}

\usepackage{subcaption}

\newcommand{\jctc}{J. Chem. Theor. Comput.}
\newcommand{\jpc}{J. Phys. Chem.}
\newcommand{\jcp}{J. Chem. Phys.}

\begin{document}

\title{The structure of 1,3-butadiene clusters}

\subtitle{Benchmarking the density-functional based tight-binding method and finite temperature properties}


\author{J. Douady         \and
        A. Simon    \and
        M. Rapacioli \and
        F. Calvo \and
        E. Yurtsever \and
        A. Tekin
}

\authorrunning{J. Douady et al.} 

\institute{J. Douady \at
              Normandie Univ, ENSICAEN, UNICAEN, CEA, CNRS, CIMAP, UMR6252, 14000 Caen, France\\
              Tel.: +33-2-31-45-25-77\\
              Fax: +33-2-31-45-25-57\\
              \email{julie.douady@ensicaen.fr}           
           \and
           A. Simon and M. Rapacioli \at
              Laboratoire de Chimie et Physique Quantiques LCPQ/IRSAMC, UMR5626, Universit\'e de Toulouse (UPS) and CNRS, Toulouse, France
        \and
        F. Calvo \at
        University Grenoble Alpes, CNRS, LiPhy, 38000 Grenoble, France
        \and
        E. Yurtsever \at
        Ko\c c University, Chemistry Department, Rumeli Feneri Yolu, Sariyer 34450, Istanbul, Turkey
        \and
        A. Tekin \at
        Informatics Institute, Istanbul Technical University, 34469 Maslak, Istanbul, Turkey
}

\date{Received: date / Accepted: date}

\maketitle

\begin{abstract}
Molecular clusters of 1,3-butadiene were theoretically investigated using a variety of approaches, encompassing classical force fields and different quantum chemical (QC) methods, as well as density-functional based tight-binding (DFTB) in its self-consistent-charge (SCC) version. Upon suitable reparametrization, SCC-DFTB reproduces the energy difference and torsional barrier of the trans and gauche conformers of the 1,3-butadiene monomer predicted at the QC level. Clusters of pure trans and gauche conformers containing up to 20 monomers were studied separately, their energy landscapes being explored using the force fields, then locally reoptimized using DFT or SCC-DFTB. The all-trans clusters are generally found to be lower in energy and produce well-ordered structures in which the planar molecules are arranged according to a herringbone motif. Clusters of molecules in the gauche configuration are comparatively much more isotropic. Mixed clusters containing a single gauche molecule were also studied and found to keep the herringbone motif, the gauche impurity usually residing outside. In those clusters, the strain exerted by the cluster on the gauche molecule leads to significant geometrical distortion of the dihedral angle already at zero temperature. Finally, the finite temperature properties were addressed at the force field level, and the results indicate that the more ordered all-trans clusters are also prone to sharper melting mechanisms.
\end{abstract}

\section{Introduction}

In the recent years, chemically complex carbon structures containing aromatic cycles have been evidenced in Space. They include fullerenes \cite{cami_detection_2010,Sellgren_2010} but also polycyclic aromatic hydrocarbons (PAHs), which are believed to represent about 10--20\% of the entire carbon budget \cite{joblin1992,draine2003}.

These observations have motivated theoretical and  laboratory studies aiming to explain how such a complexity can arise from smaller building blocks under the harsh astrophysical conditions. In particular, high-energy  collision experiments of hydrocarbon clusters have been performed to mimic the energetic processes undergone by carbon matter in shockwave regions \cite{gatchell_knockout_2016}. Assisted by molecular dynamics (MD) simulations employing the AIREBO potential, experiments showed that the  collision of 1,3-butadiene (C$_4$H$_6$) clusters with Ar$^+$ (3 keV) particles led to the formation of large (C$_4$H$_6$)$_n^+$ clusters and C$_m$H$_x^+$ molecular ions  ($n=5$--9) which could be cyclic or linear \cite{gatchell_2017}. 

The strong reactivity experienced by such clusters is not really surprising, since 1,3-butadiene is known for being a highly reactive molecule prone to polymerisation in the condensed phase \cite{buta_polym}. A detailed picture of the underlying atomic mechanisms leading to the observed products would in principle require potential energy surfaces that are not only reactive, but also correctly account for charge migration and polarization effects. In the case of (hydro)carbon clusters, the importance of correctly describing the ground and even low-lying electronic states in the context of high-energy fragmentation has been emphasized earlier \cite{Montagnon2007}.

Here we describe a first step towards this goal by employing the self-consistent charge density-functional theory tight-binding (SCC-DFTB) method \cite{scc-dftb}. This approach explicitly considers valence electrons but at a very reduced computational cost compared to other popular methods such as density-functional theory (DFT).

The SCC-DFTB approach was recently used to simulate the dissociation of PAHs at high internal energy by means of MD trajectories extending over hundreds of picoseconds and it achieved a very good agreement with mass spectroscopic measurements \cite{simon_dissociation_2017,simon_dissociation_2018,rapacioli_atomic_2018}. It thus seems ideal as well to interpret the collision experiments on butadiene clusters, and notably to address the possible importance of non-statistical knock-out mechanisms suggested by Gatchell and co-workers \cite{gatchell_2017}.  



Interestingly, despite its apparent chemical simplicity, the properties of 1,3-butadiene in the gas and condensed phases are all but well known. Obviously the intermediate case of 1,3-butadiene clusters is even less documented.
The existence of a non planar but stable gauche conformer has been experimentally evidenced only recently \cite{gauche_buta}, the equilibrium torsion angle of 34$^\circ$ being well reproduced by quantum chemical calculations. The isomerization pathway between the trans and gauche conformers had been theoretically studied earlier \cite{butarot_theo_2001}, as well as the influence of confinement on the associated barrier \cite{santiso_remarkable_2008}.

In addition to ground state studies, the electronic excited states of trans 1,3-butadiene were determined using {\em ab initio} and TDDFT methods \cite{lehtonen_coupled-cluster_2009,manna_taming_2020}. Complexes of 1,3-butadiene with rhodium and its dimer were investigated using DFT and multireference methods \cite{buta_Rh_1998}.


In the condensed phase, 1,3-butadiene has no known crystal phase. Likewise we are not aware of any structural characterization, even indirect, for finite assemblies of such molecules. The present work aims to shed some light onto the possible structures exhibited by 1,3-butadiene molecules, and to assess the SCC-DFTB method for future investigations of their reactivity towards irradiation by an impinging high-energy ion.

The workflow adopted in the present work involves different computational methods each suited to a particular goal. Classical force fields are employed to broadly sample the energy landscapes for clusters containing up to 20 monomers, imposing trans or gauche conformers. Putative lowest-energy structures are then reoptimized using both the SCC-DFTB method as well as quantum chemical methods. In doing so, we compare the energetic and structural properties and notably validate the former approach.

The article is organized as follows. The next section describes the various methods employed to model 1,3-butadiene clusters and to assess their energetic, structural but also finite temperature properties. The results are presented and discussed in Sec.~\ref{sec:res} before some concluding remarks are finally given in Sec.~\ref{sec:ccl}.

\section{Methods}

Our objective in this paper is to assess the validity and accuracy of the SCC-DFTB method to model 1,3-butadiene and its clusters. We used a variety of complementary approaches to explore the potential energy landscapes and identify likely global minima, and to refine them at the quantum chemistry level.

\subsection{Molecular mechanics}\label{subsec:mm}

A computationally inexpensive method was first chosen to exhaustively sample the structures of 1,3-butadiene clusters, namely molecular mechanics with the Amber {\em ff99} force field \cite{amber}. Two sets of parameters were employed to describe separately the trans and gauche conformers of 1,3-butadiene, with appropriate charges on all atoms taken from the DFT minima and the restrained electrostatic potential (RESP) method. By default, the gauche conformer is not stable when described with Amber {\em ff99}, because the dihedral C-C-C-C term is minimum for a vanishing torsion angle. However, the corresponding cis conformation is not a true stable conformation \cite{butarot_theo_2001} and the dihedral component of the force field had to be adjusted manually to stabilize the gauche conformation by imposing the correct torsion angle $\varphi_0$ of 35.8$^\circ$. The other two torsion angles C-C-C-H and H-C-C-H were not affected.

Because two separate sets of parameters are used to model trans and gauche conformations, the force field cannot be used to describe the rearrangement between them, or the intermediate transition states. However, it is still possible to model mixed clusters containing finite numbers of each conformation, using the appropriate parameters for each type and to the price of some minimum book keeping.

\subsection{Density-functional based tight-binding}

The force field described in subsection \ref{subsec:mm} 
provides an inexpensive tool to generate structures and sample configurational space. However it cannot account for isomerization and reactivity that take place under irradiation by multiply charged ions. We have thus also considered a more realistic description based on density-functional based tight-binding (DFTB) level, an approximated DFT scheme whose efficiency relies on the use of an atomic minimal valence basis and DFT-parameterized integrals  \cite{dftb1,dftb2,scc-dftb,dftb_rev,rev_apx}. In the present work, we have used the self-consistent charge (SCC)-DFTB scheme\cite{scc-dftb}, derived from DFT through a second order Taylor expansion of the energy functional. The second order contribution to the SCC-DFTB energy is expressed as a function of a diatomic matrix $\gamma_{\alpha\beta}$ and atomic charges $\{q_\alpha\}$ 
\begin{equation} 
E^{2nd}=\frac{1}{2} \sum_{\alpha\beta} \gamma_{\alpha\beta}(R_{\alpha\beta}) q_\alpha q_\beta,
\end{equation}
where $R_{\alpha\beta}$ is the distance between the atoms $\alpha$ and $\beta$.
In the present work, we have used the mio parametrisation set \cite{scc-dftb} combined with an empirical dispersion correction \cite{dispdftb}. We have also used  the atomic charge definition known as {\it Charge Model 3} \cite{Li_CM3,DFTB_CM3}, and a bond correction parameter $D_{\rm CH}=0.096$ which was  previously shown to improve the binding energies of molecular  clusters \cite{SimonJCP2013,DFTBreview}.

The non-zero torsion angle of the gauche isomer could not be recovered with the original SCC-DFTB potential, even when trying other sets of parameters (matsci), different types of dispersion corrections \cite{DFTB_CM3} or other CM3 bond correction parameters. In the present work, we propose to use the $\gamma$ matrix correction as introduced by Gauss {\it et al.} to improve hydrogen bonds \cite{Gaus2011tw}. In practice, only the intramolecular $\gamma$ matrix elements (i.e. $\alpha$ and $\beta$  belongs to the same molecular unit) have been modified. This correction involves an arbitrary  parameter $\zeta$ for which different values were attempted, a value of 3.0 being eventually selected  to recover reasonable structural properties, as will be discussed in section \ref{secmono}.
All these corrections can be seen as a new set of parameters and will simply be refered to as SCC-DFTB in the following, whereas the initial parameterization (mio parameters and dispersion correction) will be referred to as SCC-DFTB(old).

\subsection{Quantum chemical methods}

The accuracy of SCC-DFTB was notably assessed by performing additional calculations of the structures and energetics based on density-functional theory and using the M06-2X functional, which accounts for dispersion interactions, and the 6-31G(d,p) basis set. Basis set superposition errors were corrected by the counterpoise method, and the zero-point energy was determined in the harmonic limit from the vibrational frequencies. The Gaussian09 software package was used for these calculations \cite{g09}.

For the monomers and dimers, additional post-Hartree-Fock calculations were carried out to assess the SCC-DFTB method in its ability to describe both 
intramolecular and intermolecular components of energy. More precisely, regular MP2 and its spin-component scaled (SCS) variant were used. For the latter method, standard scaling factors of 1.2 for the singlet and 1/3 for the triplet states were used \cite{grimme03}. Coupled cluster calculations with single, double and perturbative triple excitations [CCSD(T)] were also carried out. All post-HF calculations used the basis sets aug-cc-pVDZ and aug-cc-pVTZ, and were carried out using the Molpro 2010.1 software package \cite{molpro}.

\subsection{Exploration of potential energy surfaces}

The force field was used to generate putative global minima for 1,3-butadiene clusters containing up to 20 molecules. The potential energy surfaces were sampled using the well established replica-exchange molecular dynamics (REMD) method, followed by systematic local optimizations or quenches. For these trajectories a ladder of 32 temperatures geometrically distributed between 15 and 200~K was employed, and successive simulations of 1~ns (per replica) following 1~ns of equilibration were performed, configurations being saved on the fly every 10~ps for subsequent local optimization. The REMD trajectories were repeated if a new global minimum was found, and eventually the exploration stopped once no new putative such minimum was found twice in a row.

Besides generating candidate structures, the molecular mechanics framework was also employed to give an estimate of the thermal stability of butadiene clusters in canonical equilibrium. The REMD trajectories were processed using a standard weighted histogram analysis method \cite{wham} to determine the basic thermodynamic properties such as internal energy $U(T)$ and heat capacity $C_v(T)=\partial U/\partial T$. We also considered the structural indicator of the gyration radius $R_g^2$, which roughly measures the spatial extension of the cluster and is usually expected to be larger in disordered or liquid states compared to the crystal state. In addition, we also exploited the quenching analysis performed during the optimization stage by recording the number and variety of isomers $\{ \alpha\}$ visited by the system as a function of temperature. For each isomer $\alpha$, we also counted its occurence number $n_\alpha$, or probability $p_\alpha$ after suitable normalization. This inherent structures (IS) analysis in turn provides the information entropy $S_{\rm IS}(T)$ associated with the diversity of isomers as
\begin{equation}
S_{\rm IS}(T)=-\sum_\alpha p_\alpha \ln p_\alpha,
\label{eq:IS}
\end{equation}
The inherent structure entropy is strictly zero if and only if a single conformer is visited by the system, otherwise it will have a finite positive value all the greater than fluxionality is high.

Only clusters with 10 molecules were selected for this thermodynamical analysis. Since the force field does not correctly describe the rearrangement between trans and gauche conformers, the clusters were assumed to have molecules either all in the trans conformer, all in the gauche conformer, or 9 trans molecules with one gauche molecule.

\section{Results}
\label{sec:res}

\subsection{1,3 butadiene monomer}
\label{secmono}

\begin{figure}[htb]
    \centering
    \includegraphics[width=8cm]{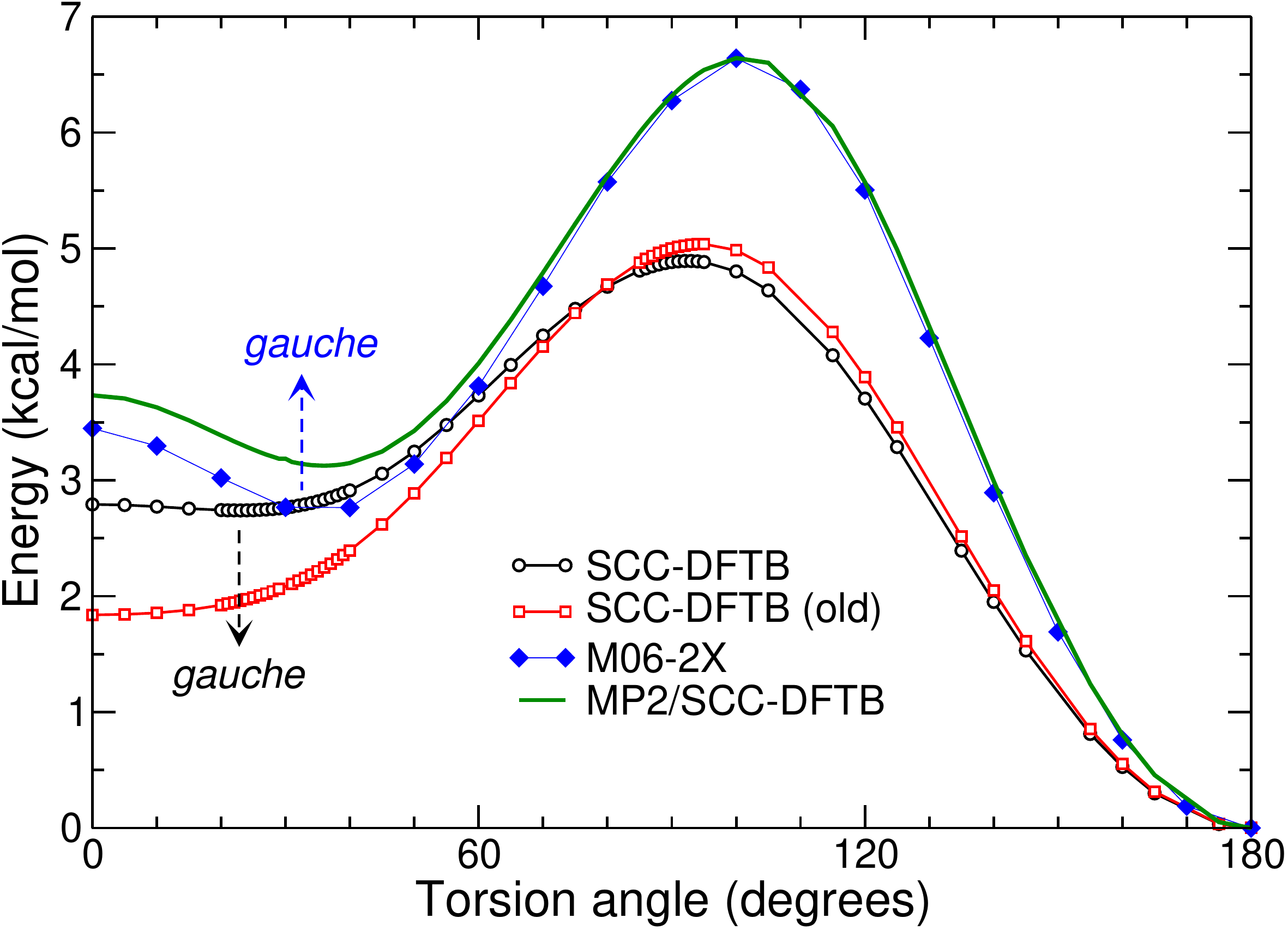}
    \caption{Energetic profile of 1,3-butadiene along the dihedral angle of the skeletal carbon atoms, obtained from the DFT (M06-2X) and SCC-DFTB methods, and MP2 single point energies from the SCC-DFTB geometries with new parametrization. The arrows locate the positions of the gauche conformation for the DFT (upward) and SCC-DFTB (downward) methods. See the text for details.}
    \label{fig:monomer}
\end{figure}

Table 1 reports experimental and theoretical data about the structural and energetic properties  of the 1,3-butadiene monomer. The entire isomerization profile obtained with several methods is depicted in Fig.~\ref{fig:monomer}.
Two planar structures can be singled out, corresponding to the trans and cis conformers, with the carbon backbone forming dihedral angles of 0 and 180$^\circ$, respectively. However, the cis geometry is not stable owing to the repulsion between facing hydrogens, and a distorsion to either of two equivalent stable gauche conformers occurs, with a dihedral angle close to $\pm$35$^\circ$ depending on the method. The trans conformer remains more stable than the gauche conformer by a few kcal/mol.

With its old parameters set, the SCC-DFTB method confirms the trans conformer to be the most stable, but does not accurately capture the gauche conformer, the optimal torsion angle being very small in magnitude and closer to the (experimentally unstable) cis structure. The energy difference between the two minima is also significantly underestimated with respect to quantum chemistry and available experimental data (Table 1).

Reparametrization improves the energetic profile to a major extent, the gauche conformer being now found near 23$^\circ$ at a relative energy of 2.73~kcal/mol in very good agreement with the DFT prediction. Single point calculation at the MP2 level on the SCC-DFTB geometry obtained with this new parametrization also matches the isomerization profile predicted by DFT, the main residual discrepancy being at the cis saddle geometry separating the two equivalent gauche conformers, where SCC-DFTB slightly underestimates the DFT result.

%
The energy barrier $E^{\dagger}$ separating the gauche conformer from the trans state is located 4.89~kcal/mol, in good agreement with experimental data (4.66--5.93 kcal/mol \cite{Carreira1975,Engeln1992}), but smaller than quantum chemistry calculations (5.98--7.54 kcal/mol). For sake of completeness, we also assessed the performance of the AIREBO potential for the 1,3-butadiene monomer. The gauche conformer is found to be stable with an angle of 18$^\circ$, which is surprisingly not too bad considering that torsions were parametrized for CH$_3$ groups only for this potential \cite{AIREBO}. However, the energy difference between the gauche and trans conformers is only 1.57~kcal/mol, in poor agreement with other reference data.


\begin{table}[htb]
    \centering
    \begin{tabular}{c c c c c c}
     \hline
    Model & $\Delta E_{\rm G}$ & $\theta_{\rm G}$ & $E^{\dagger}$ & $\nu$ & Reference\\
    & (kcal/mol) & ($^\circ$) & (kcal/mol) & (cm$^{-1}$) &\\
    \hline
    \hline
    CCSD(T) & 2.90$^{e}$ &  & 6.12$^{e}$  & & \cite{Karpfen2004} \\
    CCSD(T) & 2.95$^{a}$-2.80$^{b}$ & 35 &  &  &  This work \\
    DFT-M062X & 2.73$^{c}$(2.78)$^{c}$ & 35 & 6.63$^{c}$ & 187 & This work\\
    DFT-B3LYP       & 3.97$^{d}$ &   &  7.54$^{d}$ &  & \cite{Saha2005}\\
    SCS-MP2 & 3.03$^{a}$-2.83$^{b}$ & 35 &   &  & This work \\
    MP2 & 3.01$^{a}$-2.82$^{b}$ & 35 &   &  & This work \\
    MP2 & 2.75$^{d}$ &  & 5.98$^{d}$ & & \cite{Saha2005}\\
    MP2 & 2.90$^{e}$ &  & 6.50$^{e}$  & &  \cite{Karpfen2004} \\
    & & & & & \\
    SCC-DFTB(old) & 1.84 &$<$0.1 & 5.04 &  & This work\\
    SCC-DFTB & 2.74(2.61) & 23 & 4.89 & 140 &  This work\\
    & & & & & \\
    Exp & 2.83 & 40 & 5.93 & & \cite{Engeln1992}\\
    Exp & 2.50 & & 4.66 & & \cite{Carreira1975}\\
     \hline
    \end{tabular}
    \caption{Relative energy $\Delta E_{\rm G}$ of the gauche conformer and corresponding value $\theta_{\rm G}$ of the torsion angle, barrier $E^{\dagger}$ from the trans conformer to the gauche conformer, and imaginary frequency $\nu$ at this transition state, as obtained from various theoretical methods and available experimental data. Harmonic zero-point energy corrections are given in parentheses. The exponents $a$--$e$ refer to the basis sets aug-cc-pVDZ,  aug-cc-pVTZ, 6-31G(d,p), TZVP and cc-pV5Z, respectively.}
    \label{tab:monomer}
\end{table}

\subsection{Homo and hetero dimers}

The homo and hetero dimers built from monomers in the trans (T) and gauche (G) conformers are denoted as TT, GG, and TG, respectively. Their most stable structures can be seen in the figures corresponding to larger clusters below, and their main structural and energetic properties are reported in Table \ref{tab:dimers}.

The most accurate results expected to be provided by the coupled cluster method indicate that the TG and GG dimers are higher than the (most stable) TT dimer by 2.97 and 5.75~kcal/mol, respectively. These energies are underestimated (1.83 and 3.54 kcal/mol) at the DFTB level with the old parameter set but recovered after reparametrization (2.84 and 5.75 kcal/mol). The trans isomers remain planar at all levels of calculations. Likewise, reparametrization is necessary to prevent the gauche monomers to rearrange towards the cis saddle conformers. In contrast, quantum chemical methods and the newly parametrized SCC-DFTB approach concur to show that the gauche conformer remains significantly distorted, even though the torsion angle is slightly reduced with respect to that of the monomer.


\begin{table}[htb]
    \centering
    \begin{tabular}{c c c c c }
     \hline
   Model & System & $\Delta E$ & $\theta_{1}$ & $\theta_{2}$ \\
    & & (kcal/mol) & ($^\circ$) & ($^\circ$) \\
    \hline
    \hline
      & TT & 0   & 179 & 179 \\
     M06-2X & TG &  2.84$^{c}$(3.02) & 176 & 27 \\  
     & GG & 5.58$^{c}$ (5.58$^{c}$) & 33 & 33 \\
       & & & &  \\
      & TT & 0 &  & \\
     CCSD(T) & TG & 2.93$^{a}$-2.97$^{b}$   & -- & -- \\
      & GG & 5.74$^{a}$-5.75$^{b}$   &  &  \\      
        &  & & & \\
      & TT & 0 & & \\
     SCS-MP2 & TG & 3.02$^{a}$-2.99$^{b}$   & --& -- \\
      & GG & 5.90$^{a}$-5.77$^{b}$   & &  \\ 
        & & & &  \\
      & TT & 0   & & \\
     MP2 & TG & 2.90$^{a}$-2.86$^{b}$   & --&-- \\
      & GG & 5.78$^{a}$-5.65$^{b}$   & & \\
   & & & & \\   
      & TT & 0  & 178 & 178 \\
    SCC-DFTB  & TG & 2.84 (2.71) & 178 & 15 \\
      & GG & 5.75 (5.49) & 26 & 26 \\  
    & & & & \\   
      & TT & 0 & 179  & 179\\
    SCC-DFTB(old)  & TG & 1.83 & 179 & $<$0.2 \\
      & GG  & 3.54   & 3.4 & 3.4\\  
     \hline
           \end{tabular}
    \caption{Binding energy of the TT, GG, and TG dimers of 1,3-butadiene, with harmonic zero-point energy correction included between parentheses when available. The SCC-DFTB and DFT results correspond to full optimizations, while post-HF data were obtained as single point on the optimized DFT geometries. $\theta_{1}$ and $\theta_{2}$ are the torsion angles  for the two molecules, the values for post-HF methods being the same as the DFT values. The exponents $a$, $b$ and $c$ correspond to the basis sets aug-cc-pVDZ, aug-cc-pVTZ and 6-31G(d,p), respectively.}
    \label{tab:dimers}
\end{table}




At this stage, we can conclude that the upon the new parametrization, the SCC-DFTB method correctly describes both the intramolecular and intermolecular components of the binding energy in 1,3-butadiene. We will next use it to address larger clusters.

\subsection{Pure trans clusters}

Putative global minima of 1,3-butadiene clusters containing up to 20 molecules in the trans conformation were located at the force field level and locally reoptimized using both the SCC-DFTB and DFT methods. Their binding energy $E_b$ is defined from the total energy $E_{\rm tot}$ relative to the monomer energy $E_{\rm trans}$ as 
\begin{equation}
    E_b(n)=E_{\rm tot}(n)-nE_{\rm trans},
    \label{eq:e_b}
\end{equation}
all terms in the previous expression refering to the locally optimized value for the relevant method. For the SCC-DFTB and DFT methods, and in addition to the static values, the correction due to zero-point vibrational method was also considered but in the harmonic approximation.

\begin{figure}[htb]
    \centering
    \includegraphics[width=8cm]{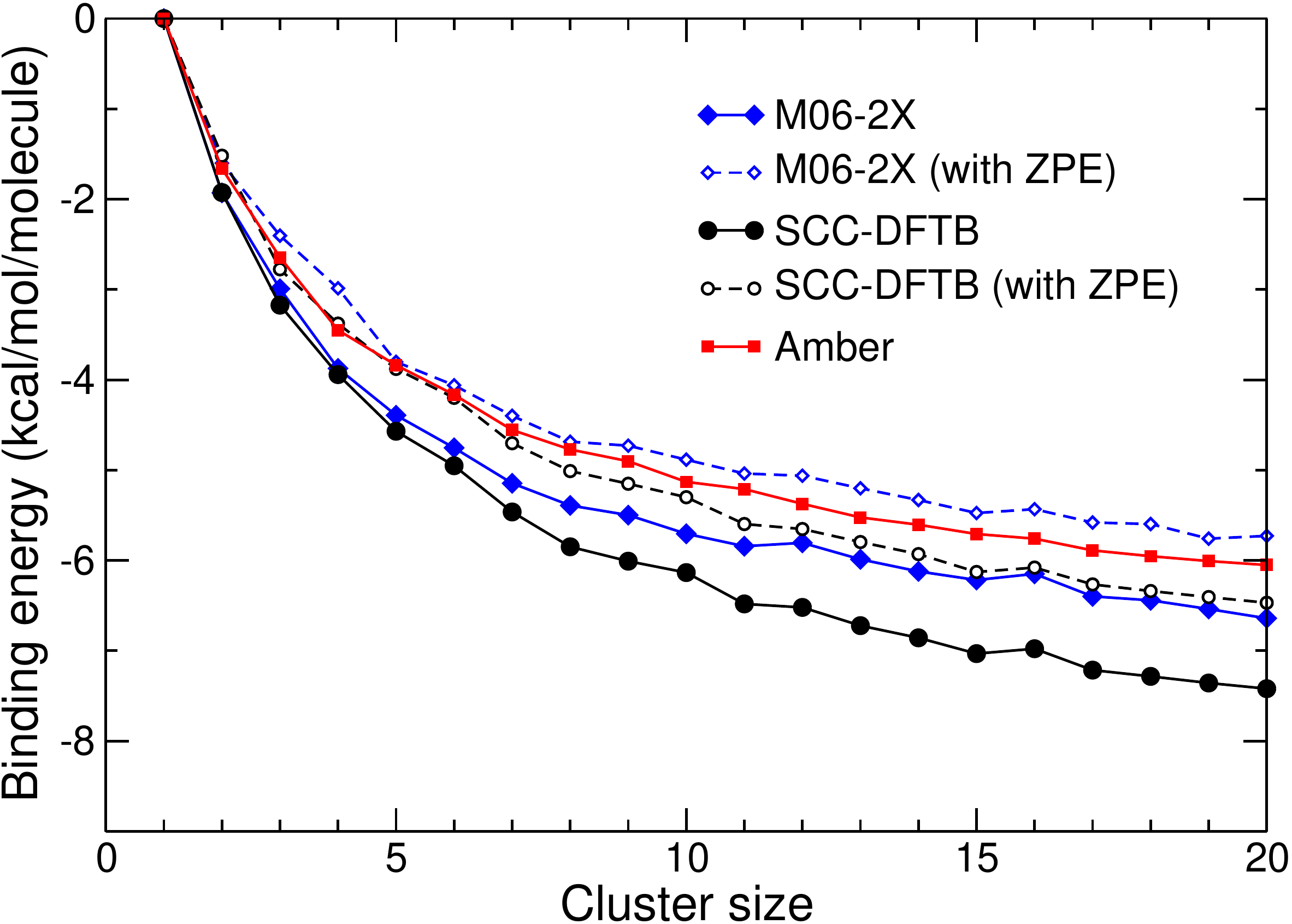}
    \caption{Intermolecular binding energy of butadiene clusters with all trans conformers, obtained from DFT and SCC-DFTB methods with and without zero-point energy corrections. The results obtained with the Amber force field to generate the putative global minima are also superimposed.}
    \label{fig:binding_trans}
\end{figure}

Fig.~\ref{fig:binding_trans} shows the variations with increasing size $n$ of the intermolecular binding energy $E_b(n)$ per molecule, as obtained for the SCC-DFTB, DFT and force field approaches.
The overall trends are rather similar for all methods, and from the static point of view the SCC-DFTB approach slightly overbinds the trans clusters relative to the DFT data, while Amber underbinds, both within 10\% magnitude. The zero-point energy correction shifts the energies to lower absolute values by approximately 1.1~kcal/mol/molecule with SCC-DFTB, 1.2~kcal/mol/molecule at the DFT level of theory. While these results are rather encouraging, they do not clearly evidence particularly stable structures for the clusters.

\begin{figure}[htb]
    \centering
    \includegraphics[width=8cm]{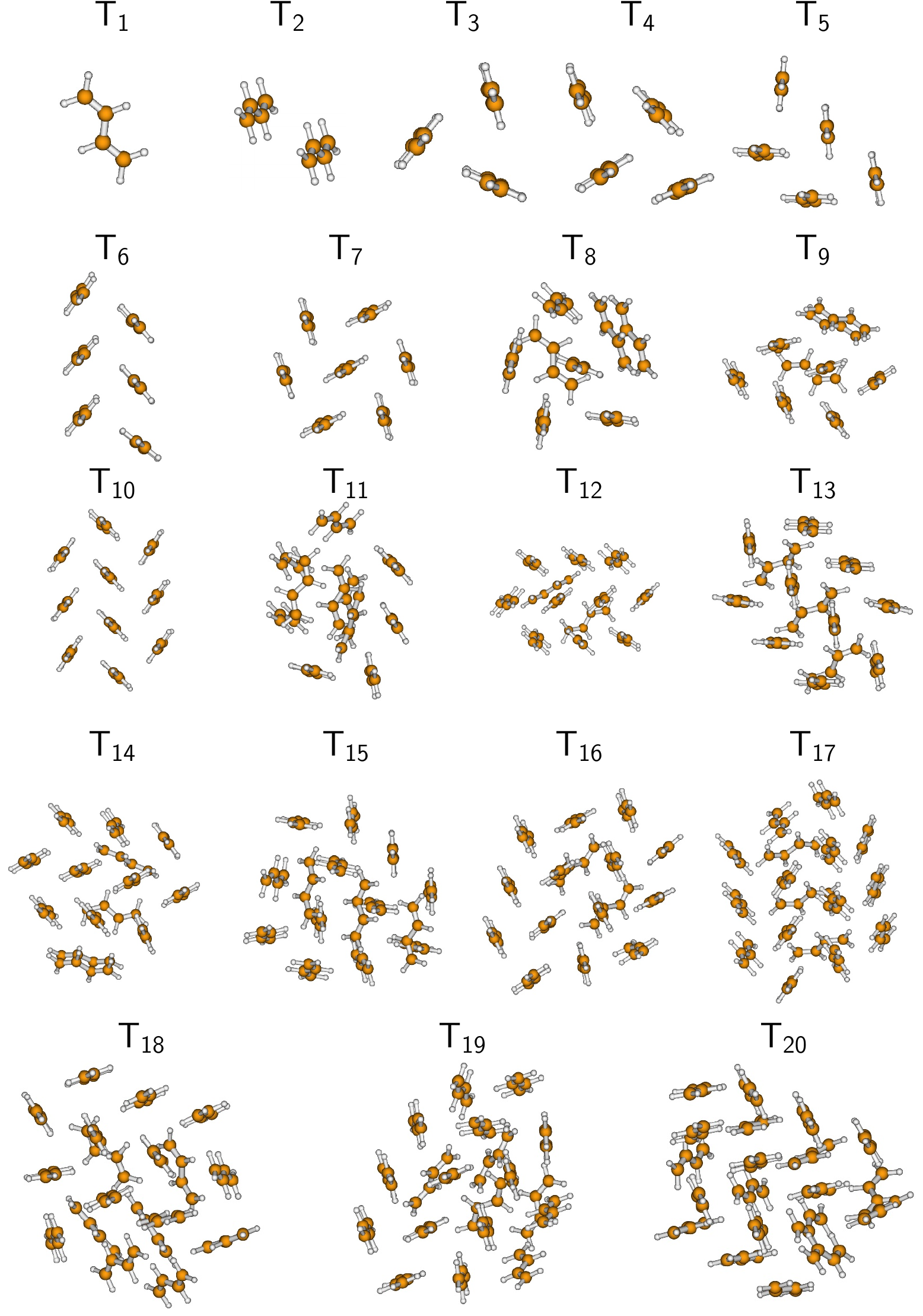}
    \caption{Lowest-energy structures of butadiene clusters with all trans conformers.}
    \label{fig:struc_trans}
\end{figure}

The usual indicators of special stability, such as the first or second difference in successive binding energy, do not reveal much additional information and no particular magic number in this size range, for the three methods employed. However, visual depiction (Fig.~\ref{fig:struc_trans}) provides interesting insight into the geometrical arrangement and growth modes of these clusters.
%
While two trans molecules tend to stack in a parallel but shifted fashion, trimers and larger clusters favor triangular arrangements evolving into the herringbone motif that is also common in other hydrocarbon clusters of comparable number of molecular units, especially 
PAHs \cite{guijarro16,takeuchi13}. Within the herringbone motif, the molecular centers of mass arrange into a common plane perpendicular to the molecular planes. Particular symmetric structures are thus obtained at $n=7$ and $n=10$, consisting of three strings of 2+3+2 and 3+4+3 molecules, respectively. Away from these sizes, it is energetically favorable to place the additional trans molecules away from the main plane, pure $\pi$-stacking in the herringbone growth mode giving rise to less spherical but also less stable structures. These additional molecules are also significantly tilted in order to maximise their dispersion interactions, a feature already visible for the 8-molecule cluster in Fig.~\ref{fig:struc_trans}. In the largest cluster at $n=20$, the butadiene molecules are also arranged into the herringbone local packing and form 2 such layers in the central part of the structure, surrounded by single strings on both side, some of the outermost molecules being again tilted to accomodate best with their local environment.

These results indicate that, in a first approach, the  structures for 1,3-butadiene clusters with all trans conformers can be predicted based on the perfect molecular crystal with herringbone motif. This could help in generating reliable geometries for arbitrarily large clusters, through appropriate truncation away from a spherical boundary.

\subsection{Pure gauche clusters}

\begin{figure}[htb]
    \centering
   \includegraphics[width=8cm]{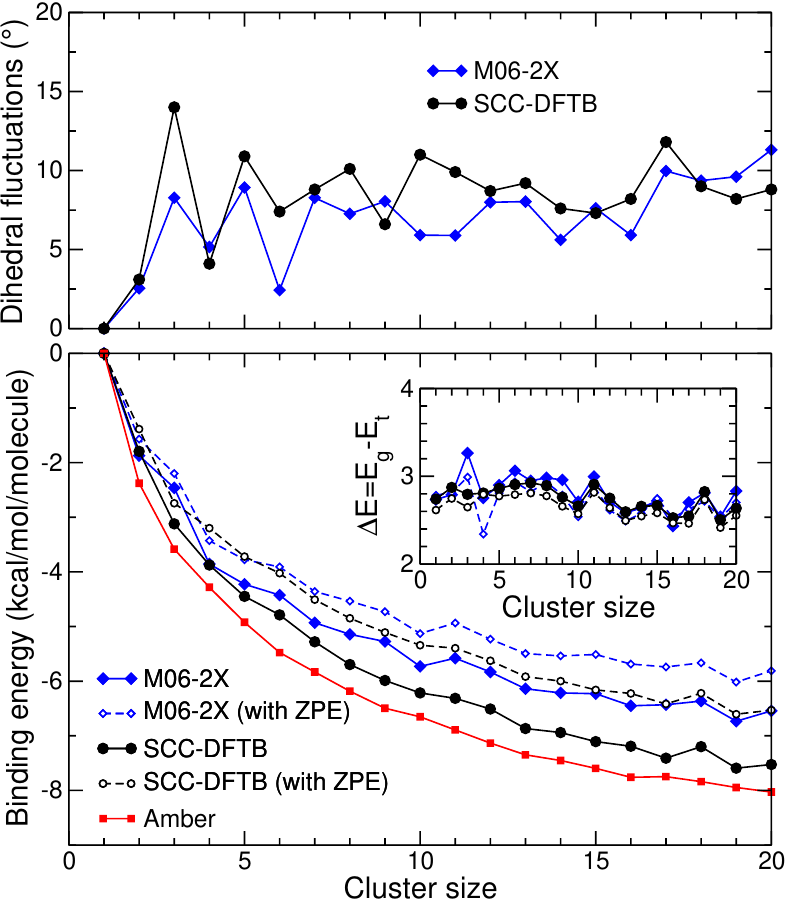}
    \caption{Energetic and geometric properties of butadiene clusters with all gauche conformers, as obtained from DFT and SCC-DFTB calculations. Upper panel: average distortion in the dihedral angle relative to the equilibrium value in the monomer. Lower panel: intermolecular binding energy, together with the value obtained with the Amber force field used to generate the structures. The inset shows the total electronic energies of the all-gauche clusters relative to the all-trans structures, in kcal/mol/molecule.}
    \label{fig:binding_gauche}
\end{figure}

1,3-butadiene clusters with all molecules in the gauche conformation were also investigated with the same methods previously employed for the pure trans clusters. Since the gauche monomer or dimer is higher in energy than the trans ones, we expect clusters containing gauche molecules to be metastable relative to their all trans counterpart. In Fig.~\ref{fig:binding_gauche}, as an inset, the total energies of the optimized gauche clusters are compared to the energies of the trans clusters for the SCC-DFTB and DFT methods. In both cases, the gauche clusters are higher by about 3~kcal/mol/molecule, a value that is similar to the monomer result (Table 1). However, this shift is also not strictly constant but varies with size within about $\pm 0.5$~kcal/mol/molecule, indicating that the molecular geometries themselves rearrange upon clustering.
Keeping in mind that these clusters are metastable, it is still possible to define their binding energy relative to the monomer units, using again Eq.~(\ref{eq:e_b}) but with the gauche energy as the monomer reference. The binding energy per molecule, shown in Fig.~\ref{fig:binding_gauche} as a function of increasing size and for the five same methods, again show comparable trends between them although now both the force field and the SCC-DFTB method overbind relative to the DFT results.The ZPE correction will also shift these binding energies per molecule towards higher values, reducing the difference between DFTB and DFT for larger clusters. We also note that, from the pure intermolecular point of view that is probed from these energetic data, the two methods that explicitly account for electronic structure predict rather similar binding energies for the all-trans and all-gauche clusters, suggesting their cohesion follows similar rules.

\begin{figure}[htb]
    \centering
   \includegraphics[width=8cm]{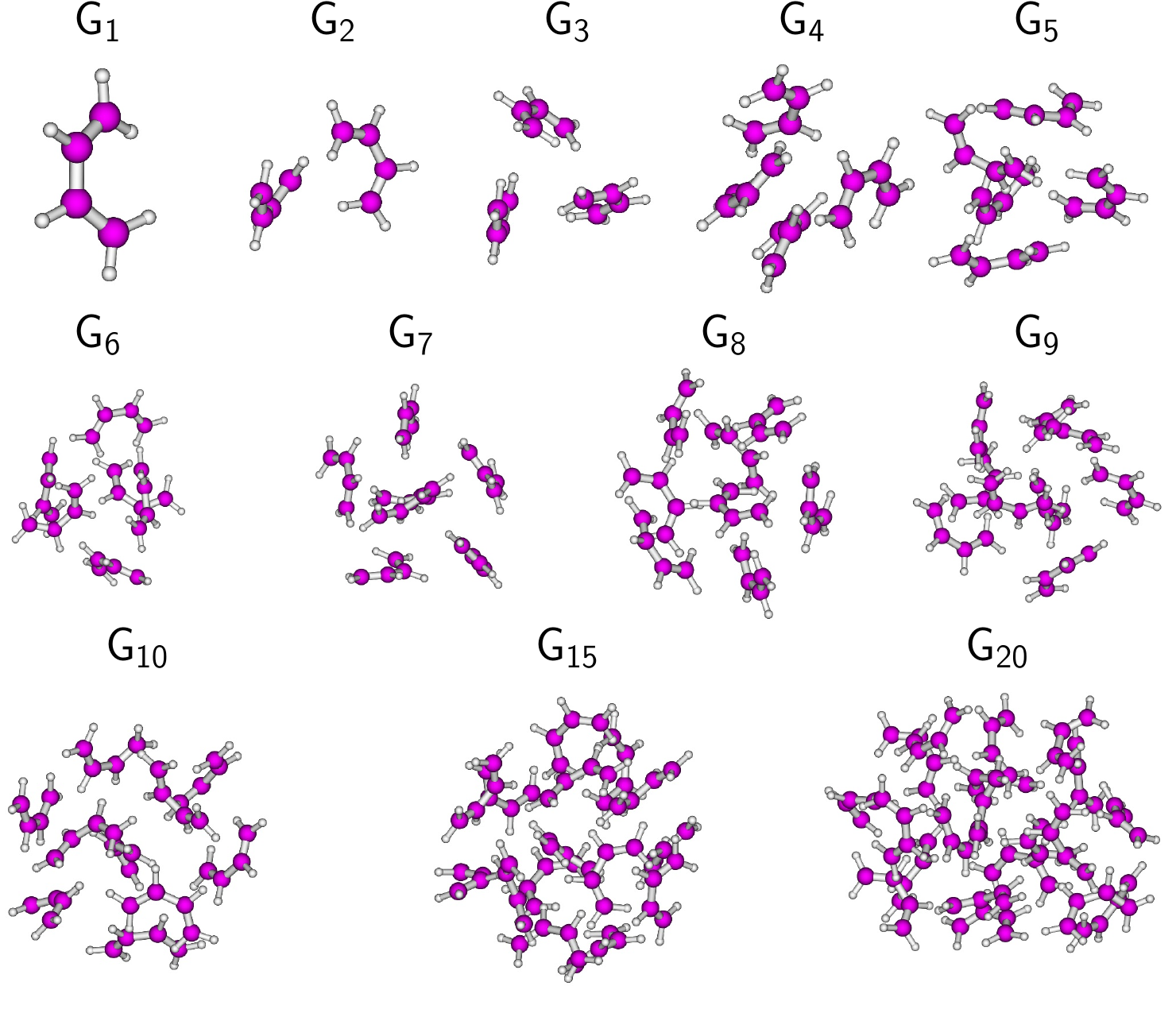}
    \caption{Selected lowest-energy structures of 1,3-butadiene clusters with all gauche conformers.}
    \label{fig:struc_gauche}
\end{figure}

A selection of the most stable structures predicted by the force field for these all-gauche clusters is shown in Fig.~\ref{fig:struc_gauche}.
%
%
Because the gauche conformer is not planar, the molecules can no longer arrange into the herringbone motif, even though the trimer looks relatively similar. The lower symmetry of the gauche conformer favors more isotropic packings than in the trans case, the molecular centers of mass adopting distorted icosahedral shapes. The molecular orientations are also in striking contrast with their counterpart in the all-trans clusters, where the herringbone pattern is associated with a specific ordering. In the all-gauche case, the molecules themselves are prone to much more significant deformations, which we quantified by determining the average deviation $\delta \theta$ in the backbone dihedral angle relative to its equilibrium value in the monomer. The variations of $\delta \theta$ with increasing cluster size, depicted in the upper part of Fig.~\ref{fig:binding_gauche}, show about 10$^\circ$ deviation in average, for both the SCC-DFTB and DFT methods, with appreciable size fluctuations. Such deformations are consistent with the rather flat potential energy surface associated with the gauche deformation accross the cis barrier, especially in the SCC-DFTB case (see Fig.~\ref{fig:monomer}).

\subsection{Mixed clusters with a single gauche impurity}

\begin{figure}
    \centering
   \includegraphics[width=8cm]{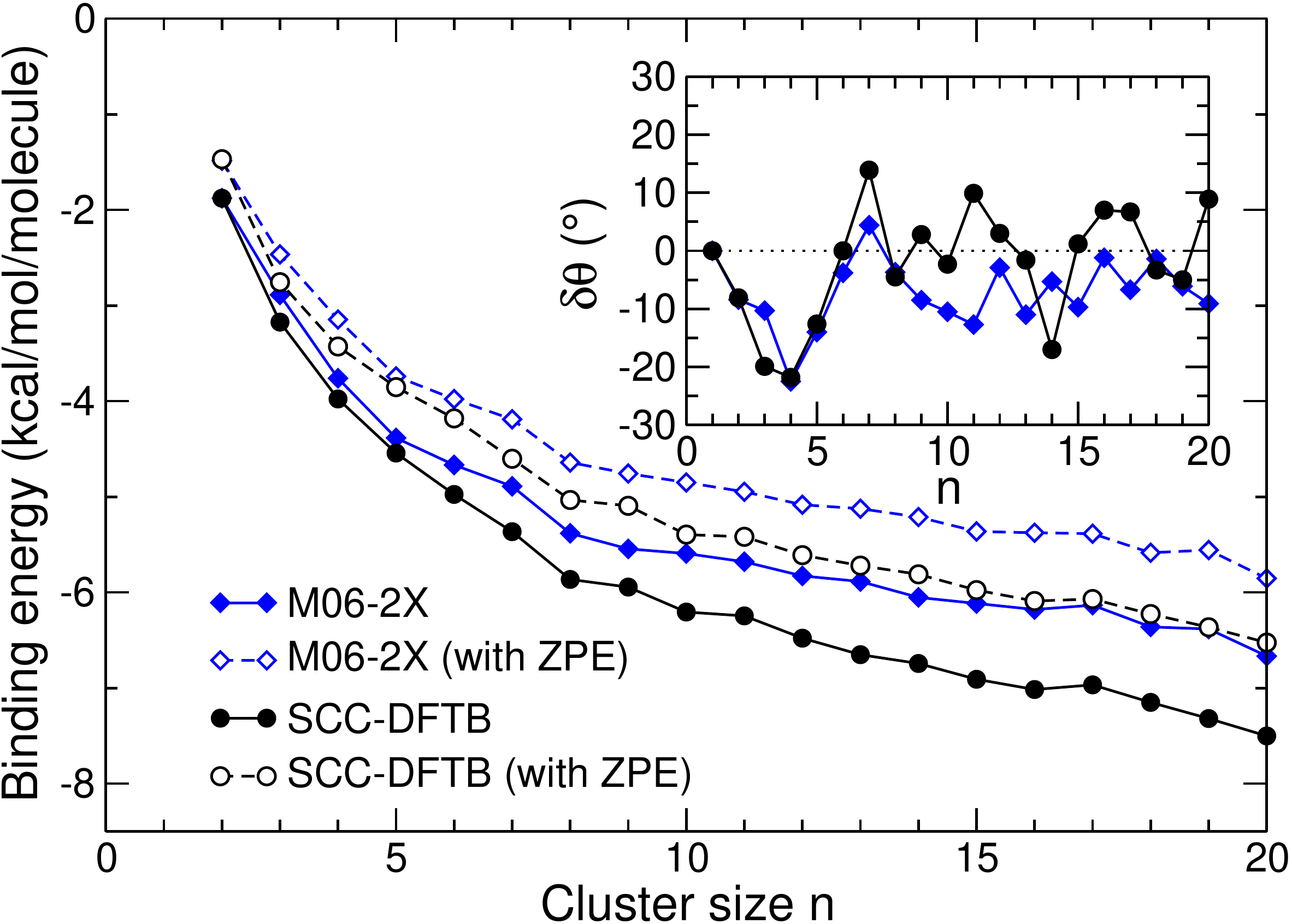}
    \caption{Binding energies of 1,3-butadiene clusters with all trans conformers but a single gauche impurity, obtained from optimizations at the DFT and SCC-DFTB level, with and without zero-point energy correction. The inset shows the deviation in the dihedral angle of the gauche conformer from its ideal value in the monomer.}
    \label{fig:binding_mixed}
\end{figure}

While pure gauche clusters of 1,3 butadiene are unlikely to be experimentally observable owing to their metastable character, the excitation brought by collision with an ion can be sufficient to isomerize one trans molecule into the gauche conformer, at least in small clusters where the excitation energy is not spread over too many degrees of freedom. We have thus considered the special case of 1,3-butadiene clusters with one gauche molecule and all other remaining molecules in the trans conformer. The binding energy for such mixed clusters, defined by analogy with Eq.~(\ref{eq:e_b}) from the total energy of the locally optimized structure after removing the binding energies of all corresponding monomers, is represented in Fig.~\ref{fig:binding_mixed} as a function of cluster size.
The trends exhibited by the SCC-DFTB method for the mixed clusters are similar as in the pure trans and gauche cases, with a binding energy slightly overestimated relative to the DFT result by about 1~kcal/mol per molecule. This result is unsurprising, as it falls right into the same order of magnitude of the corresponding data for the trans clusters, most molecules being here of the trans type.

\begin{figure}[htb]
    \centering
   \includegraphics[width=8cm]{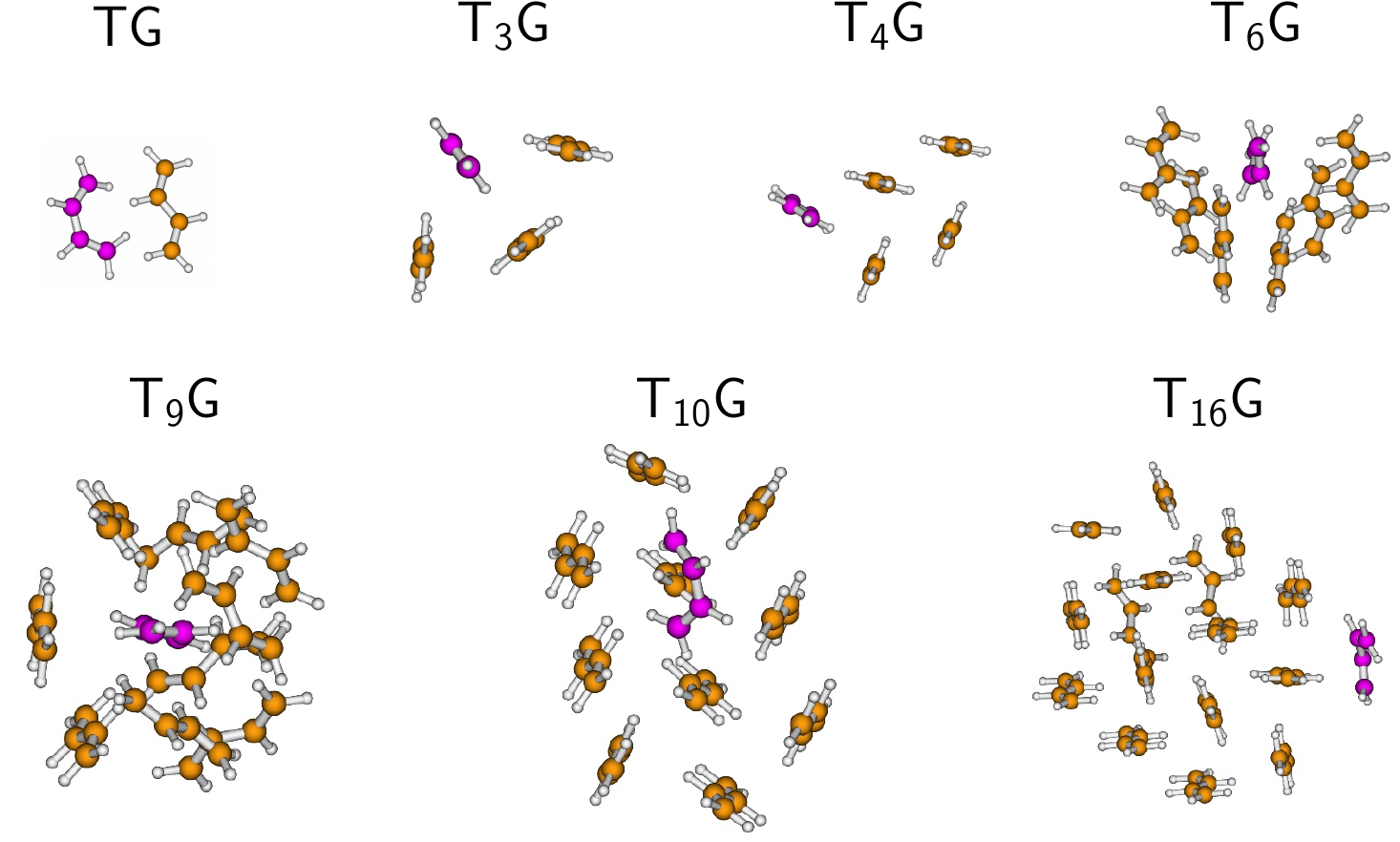}
    \caption{Selected lowest-energy structures of mixed 1,3-butadiene clusters with a single gauche conformer whose carbon atoms have been highlighted in purple color.}
    \label{fig:struc_mixed}
\end{figure}

We have also considered the possible effects of clustering on the geometrical distortion exhibited by the single gauche conformer. The absolute deviation in the backbone dihedral angle relative to its equilibrium value in the isolated monomer is represented in the inset of Fig.~\ref{fig:binding_mixed} as a function of increasing cluster size. As was the case with the pure gauche clusters, the distortion is significant and can reach $\pm 20^\circ$ with both computational methods. In particular, 
a particularly strong clustering effect can be noticed for the tetramer, for which the gauche molecule deviates by more than $-20^\circ$, making it resemble more the cis conformer that is not a locally stable configuration for the monomer. Such clustering effects can be appreciated directly from the visual depiction of these mixed clusters, for which a selection is given in Fig.~\ref{fig:struc_mixed}.
In general, the herringbone motif imposed by the trans molecules is preserved upon introduction of a gauche molecule. However this impurity plays a rather active role in dictating some occasionally strong structural changes in the entire cluster. For example, in several cases the gauche molecule has to accomodate from a surrounding of by trans molecules and distort accordingly. Here the aforementioned case of the tetramer is particularly illustrative with the nearly planar conformation exhibited by the gauche molecule, an effect that can also be noticed in the decamer to some extent. In larger clusters, the gauche impurity generally prefers staying away from the center so as not to disturb the favorable herringbone packing.

\subsection{Thermodynamical properties}

The static features exhibited by 1,3-butadiene clusters were shown in the previous section to depend sensitively on the conformation adopted by the individual molecules. In this subsection we explore the finite temperature behavior of these clusters in the range where they undergo the melting phase change. However, before addressing clusters, we have considered the individual molecule itself and its propensity for adopting the trans or gauche conformer at thermal equilibrium. Here we have used a simple harmonic approximation for the individual partition functions of each minimum, employing appropriate energy differences and vibrational frequencies at the DFT or SCC-DFTB level, as well as an additional weight of 1/2 for the trans conformer owing to its higher symmetry ($C_{2h}$ point group versus $C_2$ for the gauche conformer).

\begin{figure}[htb]
    \centering
   \includegraphics[width=8cm]{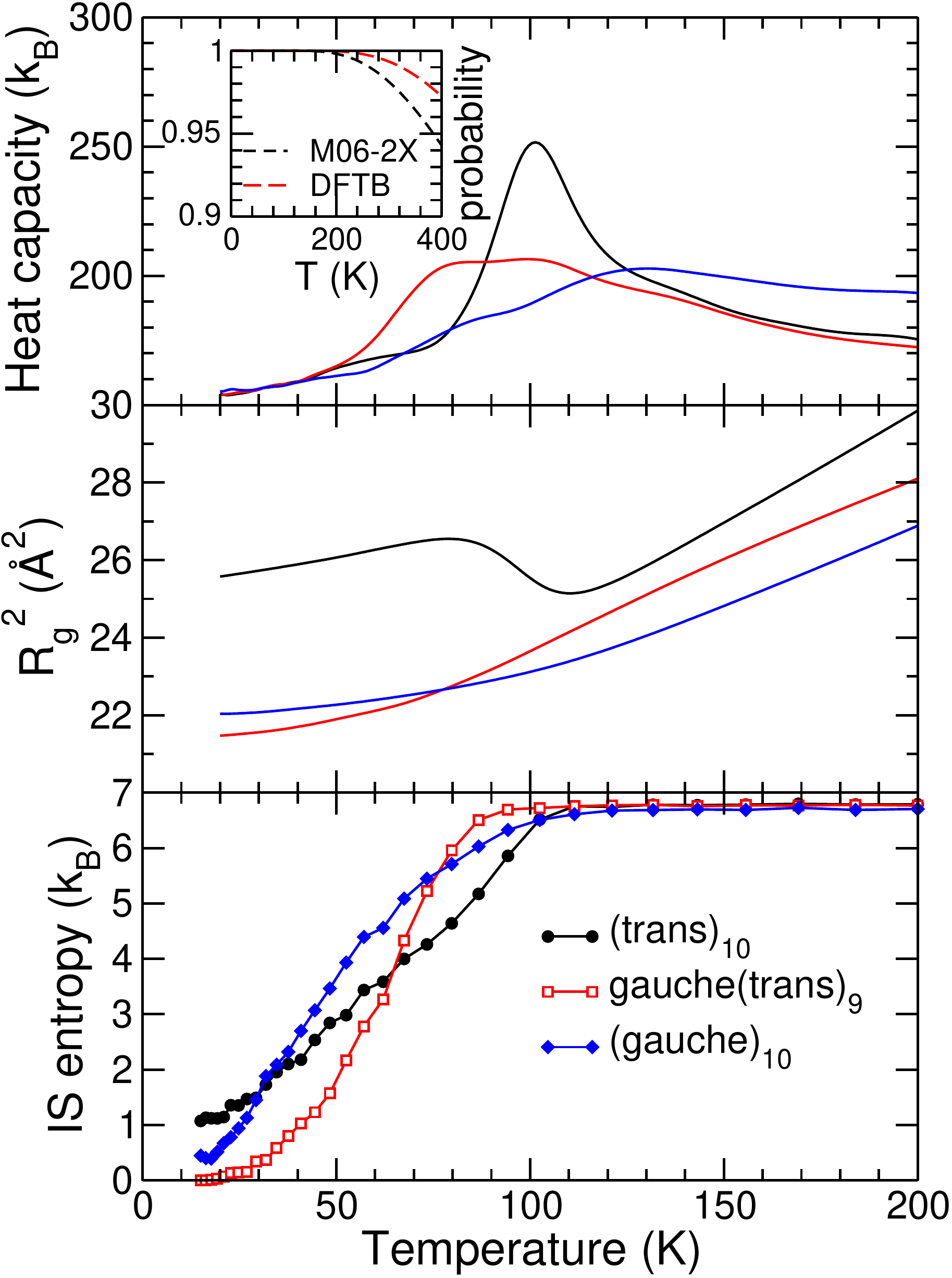}
    \caption{Finite temperature properties of 1,3-butadiene clusters with 10 molecules in the trans or gauche conformation, or assuming 9 trans molecules together with one gauche molecule, as obtained from REMD simulations with the Amber {\em ff99} force field. The upper, middle, and lower panel show the temperature variations of the heat capacity, the average square gyration radius, and the inherent structure entropy $S_{\rm IS}$, respectively. The inset in the upper panel depicts the equilibrium probability of having the monomer in the trans configuration, as obtained in the harmonic approximation with DFT or SCC-DFTB data.}
    \label{fig:thermo}
\end{figure}

The probability of finding the 1,3-butadiene molecule in the trans conformation obtained from the harmonic approximation is shown as an inset in Fig.~\ref{fig:thermo}.

The results from both the DFT and SCC-DFTB methods are consistent with one another and indicate that the trans conformer is prefered over 95\% up to 400~K, if thermal equilibrium is assumed. Therefore, clusters with molecules in the trans conformer are likely to remain as such over time up to at least room temperature. However, the gauche conformers should spontaneously interconvert into the trans form, if given enough time. This isomerization process itself depends on temperature, and as seen previously from the associated structural fluctuations, it could be strongly affected by the clustering process. Here and for simplicity we will ignore kinetic considerations and assume that the gauche molecules are trapped in this configuration.

We restrict the thermodynamical discussion to selected clusters, chosen here at the size of 10 molecules in the all-trans (T$_{10}$), all-gauche (G$_{10}$), and mixed configurations with 9 trans conformers and one gauche conformer (T$_9$G). The REMD simulations were processed to yield the canonical heat capacity $C_v(T)$, the average square gyration radius $R_g^2(T)$, and from the periodic quenches the inherent structure entropy $S_{\rm IS}(T)$ was determined.

The results obtained for the three clusters (T$_{10}$, G$_{10}$ and T$_9$G)in the temperature range $T<200$~K, depicted in Fig.~\ref{fig:thermo}, reveal some rather common trends but also sensible differences between the stable cluster T$_{10}$ and the metastable clusters containing gauche molecules.
The heat capacity shows smoothly increasing variations below 50~K and a main peak in the range 75-140~K, particularly sharp for the trans cluster but rather broad for the two other systems and extending over more than 50~K. The geometrical extension of the clusters, as monitored by the gyration radius, exhibits generally smoothly increasing variations accross the temperature range except for the pure trans cluster. In this case, a drop in $R_g^2$ occurs concomitantly with the melting peak, and is associated with the appearance of three-dimensional structures; at lower temperatures, this peculiar cluster T$_{10}$ adopts mainly structures in which the centers of mass are arranged bidimensionally in a herringbone fashion (see Fig.~\ref{fig:struc_trans}), giving rise to a particularly high gyration radius.

While the melting transition looks sharper for the pure trans cluster, fluxionality turns out to be significant also for this system already at low temperature. Its inherent structure entropy exceeds 1~$k_{\rm B}$ at 10~K, and is lower for the two clusters containing gauche conformers. Together with the variations in the gyration radius, this result indicates that in the T$_{10}$ cluster the molecules explore various bidimensional rearrangements before undergoing the melting phase change. In contrast, the other two clusters (G$_{10}$ and T$_9$G) are already essentially three-dimensional at low temperature (see Figs.~\ref{fig:struc_gauche} and \ref{fig:struc_mixed}).

Interestingly, fluxionality measured by $S_{\rm IS}$ reaches a maximum below 100~K for all three clusters, despite signatures on the thermal and geometrical indicators that are rather weak. We interpret this result as the consequence of plastic-like behavior, in which the molecules are small enough to adopt slightly different orientations while keeping their centers of mass near the equilibrium positions \cite{maillet98}.

\section{Concluding remarks}
\label{sec:ccl}

Despite the relative chemical simplicity of their building block, assemblies of 1,3-butadiene molecule deliver a rather rich phenomenology. The present article was motivated by some recent experiments in which 
clusters of 1,3-butadiene were found to undergo intracluster reactivity upon collision with a high-energy cation and produce a diversity of larger hydrocarbon molecules. A realistic modeling of the underlying processes at the atomistic scale requires chemically accurate potential energy surfaces that notably describe the monomer correctly and its ability to interconvert into the gauche conformer at the relevant excitation energies.

The strategy pursued towards this goal relies on the well established SCC-DFTB method, which is able to handle fairly large reactive systems and to address their dynamics over hundreds of picoseconds. Comparison with quantum chemical reference results indicated that some adjustments were necessary for the method to bring the gauche conformer closer to reference results.

The putative lowest-energy structures for butadiene clusters, explored by means of global optimization using a computationally inexpensive method (classical force field), reveal a dominant bidimensional growth into the herringbone motifs for the trans conformer, and a more anisotropic growth for the gauche conformer. A single gauche impurity usually alters an otherwise trans cluster especially when it lies at the center, in which case it can undergo significant deformation towards the cis conformation. Once reoptimized using density-functional theory, the SCC-DFTB method performs very satisfactorily in comparison, especially given that these van der Waals clusters are mostly bound by dispersion forces. Using the force field again, finite temperature properties were explored and the clusters found to undergo the melting phase change around 100~K.

The much higher energetic stability of the trans conformer suggests that no spontaneous interconversion to the gauche conformer should occur at thermal equilibrium, at least up to approximately 500~K. At these temperatures the clusters would not be thermally stable anyway and evaporate molecules under very short time scales. However, the energy brought by the collision with 
a charged ion is much higher, and could trigger such interconversion, as well as intracluster reactivity \cite{gatchell_knockout_2016}. Having validated here the SCC-DFTB approach for such clusters, our next effort will be dedicated to the modeling of such collisional processes and the elucidation of the particularly narrow distributions in the numbers of emitted hydrogen atoms found in the mass spectra \cite{gatchell_2017}.

\begin{acknowledgements}
This work was supported by the ANR JCJC FRAPA, grant ANR-18-CE30-0021 of the French Agence Nationale de la Recherche. Suvasthika Indrajith, Alicja Domaracka and Patrick Rousseau are gratefully acknowledged for stimulating discussions.
We gratefully acknowledge financial support from GDR EMIE 3533. We also thank C. Falvo for calculations with the AIREBO potential.
\end{acknowledgements}

%
\section*{Conflict of interest}
The authors declare that they have no conflict of interest.

\bibliographystyle{spphys}       


\end{document}